\input harvmac.tex

\def\bar{\overline}
\lref\malda{J.~ M.~ Maldacena, ``The Large N Limit of Superconformal
Field Theories and Supergravity", hep-th/9711200.}%
\lref\wholo{E.~Witten, ``Anti-De-Sitter Space and Holography'', 
hep-th/9802150.}
\lref\wther{E.~Witten, ``Anti-de Sitter Space, Thermal Phase Transition,
And Confinement In Gauge Theories'', hep-th/9803131.}
\lref\wbar{E.~Witten, ``Baryons And Branes In Anti de Sitter Space",
hep-th/9805112.}
\lref\ks{S.~Kachru, E.~ Silverstein, ``4d Conformal Field Theories
and Strings on Orbifolds", hep-th/9802183.}
\lref\dgmr{M. Douglas, G. Moore, ``$D$-Branes, Quivers, And ALE
Instantons'', hep-th/9603167.}
\lref\lnv{A.~ Lawrence, N.~ Nekrasov, C.~ Vafa,
``On Conformal Theories in Four Dimensions", hep-th/9803015.}%
\lref\bkv{M.~Bershadsky, Z.~Kakushadze, C.~Vafa, "String Expansion
as Large N Expansion of Gauge Theories," hep-th/9803076.}
\lref\dhvw{L.~Dixon, J.~A.~Harvey, C.~Vafa and E.~Witten,
``Strings on Orbifolds", \npb{\bf  261} (1985) 678; \npb {\bf 274} (1986)
285.}
\lref\dgm{M.~R.~ Douglas, B.~ R.~ Greene, D.~ R.~ Morrison, ``Orbifold
Resolution by D-Branes," Nucl.Phys.B {\bf 506} (1997) 84.}
\lref\bj{M.~Bershadsky, A.~Johansen "Large N limit of orbifold
field theories", hep-th/9803249.}
\lref\gv{R.~ Gopakumar, C.~Vafa, ``Branes and Fundamental Groups,"
hep-th/9712048.}%
\lref\wbound{E.~Witten, ``Bound States Of Strings And $p$-Branes'',
hep-th/9510135.}
\lref\go{D.~J.~ Gross, H.~ Ooguri, "Aspects of Large N Gauge
Theory Dynamics as Seen by String Theory", hep-th/9805129.}
\lref\gkp{S.~S.~ Gubser, I.~Klebanov, A.~M.~Polyakov, ``Gauge theory
correlators from non-critical string theory'', hep-th/9802109.}
\lref\malq{J.~Maldacena,``Wilson loops in large N field theory'',
hep-th/9803002.}
\lref\reyyee{S.-J. Rey and J. Yee, ``Macroscopic Strings As Heavy Quarks
In Large $N$ Gauge Theory and Anti-de Sitter Supergravity,'' hep-th/9803001.}
\lref\gk{S.~S.~ Gubser and I.~Klebanov, ``Baryons and Domain Walls in 
$\CN=1$ Superconformal Theory'', hep-th/9808075.}
\lref\akly{C. Ahn, H. Kim, B.-H. Lee, and H. S. Yang,
``$N=8$ SCFT and $M$ Theory On $AdS_4\times {\bf RP}^7$,'' hep-th/9811010.}
\lref\bsv{M.~ Bershadsky, V.~ Sadov, C.~ Vafa, ``D-Branes and Topological
Field Theories", Nucl.Phys. {\bf B463} (1996) 398.}
\lref\polchin{J.~Polchinski, S.~Chaudhuri, C.~ V.~ Johnson,
``Notes on D-Branes", hep-th/9602052.}
\lref\wlargn{E.~Witten,``Baryons in the $1/N$ Expansion'',
\npb {\bf 160}(1979) 57. }
\lref\kw{I.~Klebanov, E.~Witten,``Superconformal field theories on 
threebranes at a Calabi-Yau singularity'', hep-th/9807080.}
\lref\wsdual{E.~Witten, ``On S-duality in abelian gauge theory'', 
hep-th/9505186.}
\lref\whan{A.~Hanany, E.~Witten, ``Type IIB Superstrings, BPS Monopoles,
and Three-dimensional Gauge Dynamics,'' hep-th/9611230.}
\lref\jm{C.~V.~Johnson and R.~C.~Meyers, ``Aspects of Type IIB Theory on 
ALE Spaces,'' hep-th/9610140. }
\lref\bott{R. Bott and L. W. Tu, {\it Differential Forms In Algebraic
Topology} (Springer-Verlag, 1982).}
\lref\crocodile{A.~ Fomenko, D.~ Fuks, ``The course in homotopy topology''}
\def\IB{\relax\hbox{$\inbar\kern-.3em{\rm B}$}}
\def\IC{\relax\hbox{$\inbar\kern-.3em{\rm C}$}}
\def\ID{\relax\hbox{$\inbar\kern-.3em{\rm D}$}}
\def\IE{\relax\hbox{$\inbar\kern-.3em{\rm E}$}}
\def\IF{\relax\hbox{$\inbar\kern-.3em{\rm F}$}}
\def\IG{\relax\hbox{$\inbar\kern-.3em{\rm G}$}}
\def\IGa{\relax\hbox{${\rm I}\kern-.18em\Gamma$}}
\def\IH{\relax{\rm I\kern-.18em H}}
\def\IK{\relax{\rm I\kern-.18em K}}
\def\IL{\relax{\rm I\kern-.18em L}}
\def\IP{\relax{\rm I\kern-.18em P}}
\def\IR{\relax{\rm I\kern-.18em R}}
\def\IZ{\relax\ifmmode\mathchoice
{\hbox{\cmss Z\kern-.4em Z}}{\hbox{\cmss Z\kern-.4em Z}}
{\lower.9pt\hbox{\cmsss Z\kern-.4em Z}}
{\lower1.2pt\hbox{\cmsss Z\kern-.4em Z}}\else{\cmss Z\kern-.4em
Z}\fi}

\def\II{\relax{\rm I\kern-.18em I}}


\def\CN {{\cal N}}

\def\CW {{\cal W}}




\def\npb{{Nucl. Phys. B}}

\def\inbar{\,\vrule height1.5ex width.4pt depth0pt}
\font\cmss=cmss10 \font\cmsss=cmss10 at 7pt


\def\g{\gamma}

\def\R{\bf R}
\font\cmss=cmss10 \font\cmsss=cmss10 at 7pt
\def\Z{\relax\ifmmode\mathchoice
{\hbox{\cmss Z\kern-.4em Z}}{\hbox{\cmss Z\kern-.4em Z}}
{\lower.9pt\hbox{\cmsss Z\kern-.4em Z}}
{\lower1.2pt\hbox{\cmsss Z\kern-.4em Z}}\else{\cmss Z\kern-.4em Z}\fi}

\def\IZ{{\bf Z}}
\def\orb{{$AdS_5 \times {\bf{S^5}}/\IZ_3$}}

\Title
{\vbox{
\baselineskip12pt
\hbox{hep-th/9811048}\hbox{PUPT-1820}\hbox{IASSNS-HEP-98-93}
\hbox{ITEP-TH-30/98}
}}
{\vbox{
\centerline{Dibaryons, Strings, and Branes}
\smallskip
\centerline{ in $AdS$ Orbifold Models}
}}

\centerline{
Sergei Gukov\foot{On leave from the Institute of
Theoretical and Experimental Physics and the L.D.~Landau
Institute for Theoretical Physics}, Mukund Rangamani}
\smallskip
\centerline{\sl Department of Physics, Princeton University}
\centerline{\sl Princeton, NJ 08544, USA.}
\medskip
\centerline{and}
\medskip
\centerline{Edward Witten}                                    
\smallskip
\centerline{\sl School of Natural Sciences, Institute for Advanced Study}
\centerline{\sl Olden Lane, Princeton, NJ 08540, USA.}

\vskip 0.4cm
\def\S{{\bf S}}
\def\Z{{\bf Z}}
\centerline{\bf Abstract}
\medskip
\noindent
A generalization of the Maldacena conjecture asserts that
Type IIB string theory on \orb\ is equivalent to a certain supersymmetric
$SU(N)^3$ gauge theory with bifundamental matter.  To test this
assertion, we analyze the wrapped
branes on $\S^5/\Z_3$ and their interpretation in terms of gauge theory.
The wrapped branes are interpreted in some cases as 
baryons or dibaryons of the gauge theory and in other cases as
strings around which there is a global monodromy.  In order to successfully 
match the brane analysis with field theory, we must uncover some aspects of 
$S$-duality which are novel even in the case of four-dimensional free field 
theory.

\vskip 2cm

\newsec{Introduction}

The $AdS$/CFT correspondence proposed by Maldacena \malda\ has made it
possible to  understand many aspects of
 the large $N$ limit of conformal field
theories  in four dimensions
via Type IIB compactifications on $AdS_5 \times X$.  Here
$X$ is a compact Einstein manifold of positive curvature, 
and the conformal field
theory is formulated on the boundary of $AdS_5$.

One aspect of the correspondence is that branes wrapped on nontrivial
cycles in $X$ can be compared to states in the conformal field theory
that are nonperturbative from the point of view of the $1/N$ expansion.
In \wbar, such an analysis was made for $X=\S^5$ and ${\bf RP}^5$;
it was shown that the wrapped branes could be interpreted as soliton-like
states -- such as baryons,
strings, and domain walls --
in the large $N$ gauge theory defined on the boundary.
Some analogous results have been obtained in \gk\ for certain $\CN = 1 $
theories, and in  \akly\ for a three-dimensional field theory.

\nref\uranga{L. E. Ib\'anez, R. Rabad\'an, and A. M. Uranga,
``Anomalous $U(1)$'s In Type I And Type IIB $D=4$, $N=1$ String
Vacua,'' hep-th/9808139.}
\nref\morrison{D. R. Morrison and M. R. Plesser, ``Non-Spherical
Horizons, I,'' hep-th/9810201.}
\def\N{{\cal N}}
\def\NN{{\bf N}}
\def\B1{{\bf 1}}
\def\C{{\bf C}}
Simple examples of $X$'s with reduced supersymmetry can be constructed
as orbifolds
 \ks.  In this paper, we will consider in detail  
 the example $X=\S^5/\Z_3$, with $\N=1$ supersymmetry.
 One advantage of orbifolds is that it is comparatively easy to identify
 the boundary conformal field theory as a gauge theory.
 $AdS_5\times \S^5/\Z_3$ with $N$ units of flux
on $\S^5/\Z_3$ is the near-horizon geometry of $N$ parallel threebranes
near a $\C^3/\Z_3$ orbifold singularity.  Putting
$N$ threebranes at this orbifold gives a system with
\dgmr\
gauge group  $(U(N))^3/U(1)$ and chiral multiplets transforming
as 
\eqn\plx{(\NN,\bar\NN,\B1)\oplus (\B1,\NN,\bar \NN)\oplus (\bar \NN,\B1,\NN).}
There is also a cubic superpotential.
A conformal field theory can hardly have $U(1)$
gauge fields coupled to chiral superfields, so we are led to suspect that  
in the $AdS$ limit the $U(1)$ factors in $(U(N))^3/U(1)$
are decoupled.  (For a dynamical explanation of this decoupling
via anomalies, see \refs{\uranga,\morrison}.)  Thus, we 
suspect that the Type IIB superstring theory on $X=\S^5/\Z_3$
should be compared to an $SU(N)^3$ gauge theory with the same
chiral multiplets as in \plx.
We henceforth call this theory simply the SCFT.

This paper will be devoted to a detailed comparison of the SCFT to
Type IIB superstring theory on $AdS_5\times \S^5/\Z_3$.
A basic step in this comparison is to match up the symmetries of the
two theories.  This turns out to be surprisingly subtle; to correctly
identify the global symmetry group on the string theory side 
depends on a surprising fact, which is that under certain conditions
the operators measuring the number of $D$-strings and the number of
fundamental strings do not commute.  It is also necessary, of course,
to take into account chiral anomalies on the SCFT side.

Once the symmetries are matched,
it becomes much easier to compare the wrapped branes of the string theory
with states that are nonperturbative (with respect to $1/N$) in the
SCFT.  We identify the wrapped branes with four kinds of objects
in the SCFT, namely baryon vertices, particles, strings, and domain walls.
To be more precise, the SCFT has states that one might call baryonic,
or ``dibaryonic'' as they are built from fields charged under two
different $SU(N)$'s.  These states
 correspond to threebranes
wrapping three-cycles in $\S^5/\Z_3$ 
and strings wrapping one-cycles. There are also membranes
in $AdS_5$ formed by wrapping the threebranes on one-cycles and the fivebranes 
on
three-cycles. These membranes can end on the boundary and so look like
``strings'' in the boundary theory. There are in all 27 kinds of
such  ``gauge strings'';
it turns out that for every element of the discrete internal symmetry group of
the model, there is a string which produces that given symmetry element
as monodromy.
  This understanding of the strings
enables us also to fill in a gap in \wbar.
 Finally, fivebranes wrapping the entire manifold $X$ are interpreted
 as an external baryon vertex, 
and domain walls constructed from unwrapped threebranes have the property
that the gauge group jumps (from $SU(N)^3$ to $SU(N\pm 1)^3$) in crossing
such a wall, rather as in \wbar.

The paper is organized as follows. In section 2 we study the string theory
on the manifold \orb\ and enumerate the possible brane wrapping states,
guided by the study of non-trivial homologies of the manifold. We also
talk about the SCFT and present an analysis of the global symmetries so as
to have a complete set of quantum numbers classifying our states.  Section 3
deals with the strings and the monodromies they produce.
Details on the geometry of $\S^5/\Z_3$ are collected in the Appendix.

\newsec{The Model}

\def\IC{{\bf C}}
\def\IZ{{\bf Z}}
\subsec{The SCFT picture}

Our first task will be to analyze the symmetries and operator
content of the conformal field theory described in the introduction.

Consider Type IIB string theory on an orbifold
${\bf R^4} \times \IC^3/\Gamma$, with
$\Gamma$ being a discrete subgroup of 
the rotation group $SO(6)$ of $\IC^3={\bf R}^6$.
Upon placing $N$ D3-branes at the 
 origin of $\IC^3/\Gamma$ and taking the near horizon limit as in
\malda , we obtain Type IIB string theory on $AdS_5 \times
{\bf S^5}/\Gamma$. This construction was first analyzed in \ks , and
subsequent generalizations were discussed in \lnv . These models give simple
examples in which the AdS/CFT correspondence can be extended to
 backgrounds with
reduced supersymmetry. For example, if $\Gamma$ is contained in an $SU(3)$ 
subgroup of $SO(6)$ but not in an $SU(2)$, then the model has 
$\CN=1$ supersymmetry in four dimensions.

The AdS/CFT correspondence relates the string theory on $AdS_5
\times {\bf S^5}/\Gamma$ to the gauge theory 
that gives a low-energy description of the system of $N$ D3-branes
at the orbifold singularity $\IC^3/\Gamma$.  That latter gauge theory can
be identified by familiar orbifold methods \dgmr, \dgm, \jm.

In this paper we will focus on a simple special case:
$\Gamma$ $=$ $\IZ_3$ with the $\Gamma$ action on the coordinates 
$z_i$ of $\IC^3$ being generated by
\eqn\projection{z_i \rightarrow \exp(2 \pi i /3)  z_i.}
We will consider a system of $N$ D3-branes on $\C^3/\Gamma$,
which one can consider as coming from $3N$ such branes on the covering
space $\C^3$.  Going to the near horizon $AdS_5\times \S^5/\Z_3$ geometry,
there are $N$ units of fiveform flux on $\S^5/\Z_3$:
\eqn\flux{ \int _{{\bf S^5}/\IZ_3 } {G_{(5)} \over 2 \pi} = N. }
On the covering space $\S^5$ of $\S^5/\Z_3$, the number of flux quanta is $3N$.
The subgroup of $SO(6)$ that commutes with $\Gamma$ is $H=U(3)/\Z_3$,
and this is realized as a global symmetry group of the model.  The center
of $H$ acts as a $U(1)$ group of $R$-symmetries.  
\def\NN{{\bf N}}

The system of $N$ D3-branes at the orbifold singularity is governed by
a $U(N)^3$ gauge theory.  There are chiral superfields which should
be classified as a representation of
$U(N)^3\times H$.  Actually, it is useful to introduce the covering
group $H'=U(3)$ of $H$.  We have $H=H'/\Gamma$, where $\Gamma$ 
is the group of cube
roots of unity.
The chiral multiplets transform  in the  ${\bf 3}$ of $H'$ 
tensored with the representation
\eqn\matter{  (\NN, \bar{\NN}, 1) \oplus
(1,\NN,\bar{\NN}) \oplus (\bar{\NN},1,\NN) }
of $U(N)^3$.  As explained in the introduction, we will assume that the
$U(1)$ factors of the gauge group should be dropped before comparing
to $AdS\times \S^5/\Z_3$, and that the SCFT of interest is an $SU(N)^3$
gauge theory with the chiral superfields indicated in \matter. 
It is interesting to note that if $N$ is divisible by 3, then
a central element of $H'$ that is a cube root of unity
is equivalent to a gauge transformation by an element of the center
of $SU(N)^3$.  Hence, in this case, the connected
global symmetry group of the SCFT is the group $H$ that acts geometrically
on $\C^3/\Z_3$ and is hence manifest in Type IIB superstring theory on
$AdS_5\times \S^5/\Z_3$.  However, if $N$ is not divisible by 3, then
no nontrivial element of $H'$ is equivalent to a gauge transformation,
and the connected global symmetry group of the SCFT really is the threefold
cover $H'$ of the geometrical symmetry group $H$.  
At the end of section 2.2, we will see how this comes about
in string theory on $AdS_5\times \S^5/\Z_3$.

We denote the matter superfields of the $SU(N)^3$ theory
as $U_{\mu}$, $V_{\mu}$, $W_{\mu}$ respectively, where  $\mu \in \{1,2,3\}$
labels the ${\bf 3}$ of $H'$, while $U$, $V$, and $W$ are associated
with the three summands in \matter.  
 The most
general cubic superpotential with $SU(N)^3\times H'$ symmetry is
\eqn\supot{ \CW =\g \epsilon_{\mu \nu \rho} 
U^{\mu} V^{\nu} W^{\rho}} 
with $\g$ a constant.  For the orbifold, this superpotential is actually
present with nonzero $\g$.

So far we have considered only the connected part of the global symmetry
group.  In comparing to the string theory, it will be very important
to also understand the discrete global symmetries.  

One obvious  symmetry
is a cyclic permutation of the three $SU(N)$ factors in the gauge group,
accompanied by $(U,V,W)\to (V,W,U)$.  This gives a $\Z_3$ symmetry
group, whose generator we will call $A$.

To look for more discrete symmetries, we consider 
$(U,V,W)\to (aU,bV,cW)$, where $a,b$, and $c$ are complex numbers 
of modulus one.  There is no essential loss in considering only choices
of $a,b,c$ under which the superpotential is invariant (since we have
already identified $R$-symmetries), so we assume $abc=1$.    
Absence of anomalies under $SU(N)^3$ instantons gives
$(ab)^{3N}=(bc)^{3N}=(ca)^{3N}=1$; using also $abc=1$, we get
$a^{3N}=b^{3N}=c^{3N}=1$.  Moreover, a transformation with $a^N=b^N=c^N=1$
is equivalent to a gauge transformation by an element of the center of
$SU(N)^3$.  If we set $\zeta=\exp(2\pi i/3N)$, then (modulo gauge 
transformations), the interesting choices of $a,b$, and $c$
are generated by $B:(a,b,c)=(\zeta,\zeta^{-1},1)$ and $C:(a,b,c)
=(\zeta^{-2},\zeta,\zeta)$.  One has $B^3=C^3=1$ (modulo gauge transformations)
and of course $B$ and $C$ commute. 

Now we find a very interesting detail.  Modulo gauge transformations,
$A$ and $C$ commute, but
\eqn\uxxz{AB=BAC.}
Thus, $A,B,$ and $C$ generate a nonabelian group $F$ with 27 elements.
Actually, it is somewhat imprecise to call $F$ a discrete symmetry
group; this is so if $N$ is divisible by 3, but otherwise $C$ is equivalent
modulo a gauge transformation to the element $T^N$ of $H'$,
where  $T=e^{2\pi i/3} $.
(In proving this, one must use the fact that if $N$ is not divisible
by 3, then $N^2-1$ is divisible by 3. It follows that $\exp(2\pi i((1/3N)
-N/3))$ is an integral power of $\exp(2\pi i/N)$, and so is an element of 
the center of $SU(N)$.) 
We will not incorporate this in our terminology and will
refer to $F$ as a discrete symmetry group.

The group $F$ admits the following action of $SL(2,\Z)$ by outer
automorphisms.  An element
\eqn\gicco{M=\left(\matrix{ a & b \cr c & d}\right)}
of $SL(2,\Z)$ acts by \eqn\jollygood{
A\to A^aB^b, \,B\to A^cB^d, ~~C\to C.}
(Of course, this transformation only depends on the reduction of $M$ modulo 3.)
The model is expected to have an $SL(2,\Z)$ $S$-duality symmetry, inherited
{}from the $SL(2,\Z)$ symmetry of Type IIB in ten dimensions.  
We propose that $S$-duality acts on the discrete symmetries in the way
just indicated.  This proposal will be incorporated in our proposal
for matching the SCFT with Type IIB superstrings on $AdS_5\times \S^5/\Z_3$.

The SCFT has a few other discrete symmetries that will be much
less important in the present paper and which  we note only briefly.
If $\gamma$ is real, then there is a parity symmetry $P$ in which exchange
of two factors in the gauge group is accompanied by orientation reversal
of spacetime.  There is also a
charge conjugation symmetry $C$ which exchanges two factors of the
gauge group and acts in each
factor of the gauge group as the outer automorphism that maps $\NN$
to $\bar \NN$.  
In the string theory on $AdS_5\times \S^5/\Z_3$, P corresponds to
an orientation-reversing symmetry of $\S^5/\Z_3$ combined with one of
$AdS_5$, and $C$ to the world-sheet orientation reversal $\Omega$.

\bigskip\noindent{{\it Nonperturbative Excitations In The $1/N$ Expansion}}

Now let us discuss the spectrum of the SCFT.  First, we consider
states that are perturbative from the point of view of the $1/N$ expansion.
These are states that can be built from a fixed number of elementary
excitations, independent of $N$.  If we let $n_U$ be the number
of $U$ fields minus antifields (and including, of course, the fermionic
partners of $U$), and define similarly $n_V$, $n_W$, then a simple exercise
in $SU(N)^3$ group theory shows that all gauge-invariant
excitations made from a
fixed number of quanta (independent of $N$) have $n_U=n_V=n_W$.
Hence, such excitations are invariant under $B$ and $C$.  On the
other hand, $A$ can perfectly well act nontrivially on perturbative
excitations.  Thus, the 27-element group $F$ has a $\Z_3\times \Z_3$ subgroup,
generated by $B$ and $C$, that acts trivially on states that
are perturbative in the $1/N$ expansion.

What about nonperturbative states?  One can build in this theory
a gauge-invariant operator, nonperturbative with respect to the $1/N$
expansion, of the form $\epsilon^{i_1\dots i_N}\epsilon_{j_1\dots j_N}
U^{j_i}_{i_1}\dots U^{j_n}_{i_N}$.  We schematically denote this
state as $U^N$.  One can build analogous states $V^N$ and $W^N$.  We will
call these states baryons, or dibaryons.  $B$ and $C$ act nontrivially
on dibaryons.

In addition to these baryonic states (which are somewhat analogous
to the Pfaffian states considered in \wbar\ in the case of $SO(2n)$ gauge
theory), one can consider baryon vertices connecting external charges.
For example, $N$ external charges in, say, the first $SU(N)$ factor
in the gauge group can be combined to a gauge-invariant state using
the antisymmetric tensor $\epsilon_{i_1i_2\dots i_N}$.  We will want to
describe such a baryon vertex in terms of $AdS_5$.

At first sight, it may seem that there are three kinds of baryon vertex
to consider -- as one could have external charges in any of the three
$SU(N)$ factors in the gauge group.  However, modulo emission and absorption
of the baryonic particles $U^N$, $V^N$, and $W^N$, the three types of
baryon vertex are equivalent.  For instance, since the $U$ field
transforms as $(\NN,\bar\NN,{\bf 1})$ under $SU(N)^3$, a baryon
vertex in the second $SU(N)$ plus a $U^N$ state is equivalent to a baryon
vertex in the first $SU(N)$.

\subsec{String theory on \orb}

Type IIB on $AdS_5\times \S^5/\Z_3$ has obvious symmetries that
come from geometrical symmetries of this manifold, namely the 
Anti-de Sitter symmetry group and the $H=U(3)/\Z_3$ symmetry group
of $\S^5/\Z_3$.
To identify additional symmetries of Type IIB superstring theory on
$AdS_5\times \S^5/\Z_3$,
we must look at the possibilities of brane wrapping.
The nontrivial integral homology groups of $\S^5/\Z_3$ are
\eqn\homology{\eqalign { 
& H_0({\bf S^5}/\IZ_3 , \IZ) = H_5({\bf S^5}/\IZ_3 , \IZ) = \IZ 
\cr 
& H_1({\bf S^5}/\IZ_3 , \IZ) = H_3({\bf S^5}/\IZ_3 , \IZ) = \IZ_3  .
\cr}} 
A generator of $H_1(\S^5/\Z_3)$ is a linearly embedded $\S^1/\Z_3$
subspace; a generator of $H_3(\S^5/\Z_3)$ is similarly a linearly
embedded $\S^3/\Z_3$ subspace.\foot{To be precise about this,
let $\S^5$ be the subspace $|z_1|^2+|z_2|^2+|z_3|^2=1$ in $\C^3$,
with $\Z_3$ acting by $z_i\to e^{2\pi i/3}z_i$.  Then, up to a $U(3)$
transformation, $\S^1/\Z_3$ is defined by $z_2=z_3=0$, and $\S^3/\Z_3$
is defined by $z_3=0$.}

The possibilities for brane wrapping  are thus as follows:

\item{{\it (i)}}
We can make particles 
in $AdS_5$ by wrapping a $p$-brane on a $p$-cycle for
$p=1,3,5$.  (In some cases, these objects actually turn out
to be baryon vertices, connected to the boundary by strings, rather
than localized particles.)

\item{{\it (ii)}}
We can make membranes in $AdS_5$ by wrapping a $p$-brane
on a $(p-2)$-cycle for $p=3,5$.

\item{{\it (iii)}}
We can wrap fivebranes on one-cycles in $\S^5/\Z_5$ to make an object
that completely fills $AdS_5$.

\item{{\it (iv)}} Finally, $p$-branes that are not wrapped at all
on $\S^5/\Z_3$ look like $p$-branes on $AdS_5$.

Wrapped branes of type {\it (iii)} really correspond to having a different
$AdS_5$ theory, giving something that should be compared not to the 
SCFT we have described, 
but to a different (possibly nonconformal) boundary theory.
This possibly interesting direction will not be explored
in the present paper.  The unwrapped branes, type {\it (iv)}, are also
easy to dispose of.  The unwrapped onebranes are related to Wilson
and 't Hooft loops in the boundary conformal field theory, as
in \malq, \reyyee.  The unwrapped threebranes are domain walls, across
which the $SU(N)^3$ theory jumps to an $SU(N\pm 1)^3$ theory, by the same
reasoning as in 
\wbar.  We will concentrate in the present paper primarily on
wrapped branes of types {\it (i)} and {\it (ii)}.

Concerning type {\it (i)}, the  fivebranes that are entirely
wrapped on $\S^5/\Z_3$ can be interpreted
precisely as in \wbar\ in terms of baryon vertices connecting
external quarks.  To be specific, the totally wrapped $D5$-brane
is a baryon vertex connected by elementary strings
to $N$ external electric charges; modulo emission and absorption of ordinary
particles (localized
AdS excitations), there
is only one such vertex (rather than one for each factor in the gauge
group) for reasons explained at the end of section 2.1.

The other objects of type {\it (i)} are a  fundamental
string or  $D$-string wrapped on $\S^1/\Z_3$, 
and a threebrane wrapped on $\S^3/\Z_3$.
The number of such wrapped objects (of any of the three kinds)
 is conserved modulo 3, since the relevant homology
groups of $\S^5/\Z_3$ are both isomorphic to $\Z_3$.
Let $A'$ be the operator that counts wrapped fundamental strings
(on a state with $k$ such strings, the eigenvalue of $A'$ is
$\exp(2\pi i k/3)$), and similarly let $B'$ and $C'$ be the operators
that count the numbers of wrapped $D$-strings and wrapped threebranes,
respectively.  

We would like to compare the symmetry generators
called $A'$, $B'$, and $C'$ here
with the operators $A,B, $ and $C$ of the SCFT.
We note that wrapped fundamental strings can be seen in string perturbation
theory and so should correspond to perturbative objects in the $1/N$ expansion
of the SCFT.  
By contrast, the other wrapped branes are nonperturbative objects in
string perturbation theory and should be nonperturbative in the $1/N$
expansion of the boundary theory.

Hence, comparing to our analysis of the discrete symmetries
of the SCFT in section 2.1, 
we identify $A'$ with $A$.  This identification can actually
be justified directly by considering the $\C^3/\Z_3$ orbifold.
The orbifold has of course a quantum $\Z_3$ symmetry (which acts
trivially on strings in the untwisted sector and nontrivially on
twisted sectors).  For $N$ threebranes near the orbifold singularity,
the quantum $\Z_3$ symmetry becomes \dgmr\ the group of cyclic permutations
of the three $SU(N)$'s; the generator of this group is what we have called
$A$.  The twisted sector states, on which $A$ acts nontrivially, become
wrapped fundamental strings when we go to the near horizon
$AdS_5\times \S^5/\Z_3$ geometry, and this explains why $A=A'$.

Now, in section 2.1, we worked out the commutation relations of $A,B$, and $C$,
and discovered an $SL(2,\Z)$ group of outer automorphisms that intertwines 
$A$ and $B$ according to \jollygood.  In Type IIB superstring theory,
there is an $SL(2,\Z)$ $S$-duality group that intertwines in precisely the
same way the operators $A'$ and $B'$ measuring the numbers of wrapped
strings.  We thus  extend our identification of $A$ with $A'$
to identify $B$ with $ B '$.

Finally, by default, we are left to postulate that $C$ should be identified
with $C'$. (We will also give a fairly direct argument for this below.)
 Here, we must face the following puzzle.  In string theory,
it appears that the operators $A'$, $B'$, and $C'$ measuring the numbers of
wrapped branes of different kinds should all commute.  They thus
appear to generate the 27-element group $(\Z_3)^3$.  However,
in section 2.1, we learned that $A,$ $B$, and $C$ generate a
{\it nonabelian} group with 27 elements.  What is the origin of this
discrepancy?  In section 2.3 below, we will analyze this question
and show that in fact, the operators $A'$, $B'$, and $C'$ do not
commute and obey instead 
\eqn\jxu{A'B'=B'A'C'.}

Since (according to our
hypothesis) $C'=1$ in the absence of wrapped threebranes, and 
$C'=\exp(\pm 2\pi i/3)$ when a wrapped threebrane or anti-threebrane is 
present,
the concrete meaning of this statement is that although $A'$ and $B'$  commute
in the absence of a wrapped threebrane, they no longer commute in the presence
of such a brane.  Concretely, a wrapped threebrane supports
a $U(1)$ gauge field, and suitable states of this gauge field carry
$F$-string (fundamental string) and $D$-string number.  The operators
$A'$ and $B'$ become at low energies
essentially the Wilson and 't Hooft loop operators of the $U(1)$ gauge
theory.  Thus, the assertion that $A'$ and $B'$ do not commute
in the presence of the threebrane (but obey \jxu) will be justified
by establishing a novel effect in free field theory,  more specifically
in $U(1)$ gauge theory in four dimensions.

The other main subject of the rest of this paper will be the wrapped branes
of type {\it (ii)}, which give twobranes on $AdS_5$.  These objects are
of codimension two, so there can be a monodromy in going around such
an object.   Taking 0, 1, or 2 
threebranes on $\S^1/\Z_3$, and 0, 1, or 2 Dirichlet or NS
fivebranes on $\S^3/\Z_3$, with all these objects parallel to each other
on $AdS_5$, we see that there are 27 possible membranes, counting
the trivial one. 27 is the order of the global symmetry group
$F$, and this suggests that each element of $F$ is the monodromy around
one of the membranes.  If so, it is clear that $C'$, which is 
$SL(2,\Z)$-invariant, must be the monodromy around a membrane made by
wrapping a threebrane, while $A'$ and $B'$ must be the monodromies
around membranes constructed from wrapped fivebranes.   Justifying
these statements will be the goal of section 3.

\bigskip\noindent
{\it Extension Of The Global Symmetry Group}

To tie up some loose ends and further justify the identification
of the threebrane wrapping number $C'$ with $C$, we now examine
the quantization of the wrapped threebrane.   We want to see how
the symmetry group $H=U(3)/\Z_3$ of $\S^5/\Z_5$ is extended
to $H'=U(3)$, as predicted in section 2.1, when $N$ is not divisible
by 3.

A threebrane wrapped
on a particular $\S^3/\Z_3\subset \S^5/\Z_3$ is invariant under
a subgroup $U(2)/\Z_3$ of $U(3)/\Z_3$.  The space of such classical
configurations is thus a copy of $(U(3)/\Z_3)/(U_2/\Z_3)=U(3)/U(2)=
{\bf CP}^2$.  The wrapped threebrane is thus equivalent at low
energies (and large $g^2N$) to a particle moving on ${\bf CP}^2$.
Because the threebrane is electrically charged with respect to $N$ units
of five-form flux, the wave function of this particle is a section of
the line bundle ${\cal L}^N$, where ${\cal L}={\cal O}(1)$ is the usual
ample line bundle over ${\bf CP}^2$.  The holomorphic sections of
${\cal L}^N$ $-$ which give the lowest energy states of the wrapped 
threebrane $-$ transform in the $N^{th}$ symmetric tensor representation 
of $H'=U(3)$.  
This representation is faithful 
if $N$ is not divisible by three, showing, as we saw in section 2.1 from the 
point of view of the SCFT, that for $N$ not divisible by three,
$H$ is extended to $H'$.

We can be more specific about this.  Supposing that $N$ is not divisible
by 3, let $T$ be the element
$\exp(2\pi i/3)$ of $H'$.  Thus $T$ measures the ``triality'' of
an $H'$ representation.  
$T$ acts trivially on states that contain no wrapped threebranes.
(It acts trivially on perturbative string states, since it acts trivially
on the spacetime $AdS_5\times \S_5$.  Also, by quantizing
the appropriate collective coordinates, it can be seen to act trivially
on wrapped onebranes.)
But on a state with a wrapped threebrane, $T$ acts, given what we have seen
in the last paragraph, as $\exp(2\pi i N/3)$. 
This is the same as the triality or $T$ eigenvalue of the dibaryon state
$U^N$, supporting the idea that the wrapped threebrane is a dibaryon.
The relation between $C'$ and $T$ can be written $T=(C')^N$ or
equivalently if $N$ is not divisible by 3 (and hence $N^2$ is congruent
to 1 modulo 3) $C'=T^N$.  We found the same formula for $C$ in section
2.1, supporting the relation $C'=C$.

\subsec{Topologically Nontrivial 't Hooft And Wilson Lines}

It remains to explain an important detail.  As we have seen, for
string theory on $ AdS_5\times \S^5/\Z_3$ to agree with
the SCFT, it must be that in the presence of a wrapped threebrane,
the operators $A'$ and $B'$ measuring the number of wrapped fundamental
strings or $D$-strings do not commute.

On the worldvolume of the wrapped threebrane -- which for our purposes
is a copy of $\S^3/\Z_3$ -- there is a $U(1)$ gauge field $a$.  Suitable 
configurations of this gauge field, roughly with nonzero eigenvalues of Wilson
or 't Hooft loops, carry fundamental string or $D$-string charge.
Thus, our question amounts to a question about free $U(1)$ gauge
theory on $\S^3/\Z_3$.

In general, on a Type IIB threebrane of any topology, the induced fundamental
string charge is measured by the first Chern class of the $U(1)$ line bundle.
Complex line bundles on $\S^3/\Z_3$ are classified by their first Chern class
which takes values in 
\eqn\iccok{H^2(\S^3/\Z_3,\Z)=\Z_3.}
Thus, as expected, the fundamental string charge on $\S^3/\Z_3$
is $\Z_3$-valued.

A line bundle whose first Chern class is torsion admits a flat connection.
The fundamental group of $\S^3/\Z_3$ is $\Z_3$, and a flat connection is 
specified by up to isomorphism by giving its monodromy
around a circle $\alpha$ that generates this $\Z_3$.  The monodromy is of 
the form $\exp(2\pi ik/3)$, where $k=0,1,$ or 2 is the fundamental
string number.  Thus, for each line bundle, there is a minimum energy
state, associated with the flat connection on that line bundle.  For
the flat connection on the 
$k^{th}$ line bundle, the value of the Wilson line $W=\exp i\int_\alpha a$ 
is $\exp(2\pi ik/3)$, and this is the expected eigenvalue in that sector
of the operator $A$ that counts fundamental wrapped strings.  
In that sense, the fundamental string operator is related to the Wilson line.

Dually, we expect to measure the number of wrapped $D$-strings by an
't Hooft loop operator on the circle $\alpha$.  Here we will meet
a very interesting subtlety which will lead to the expected formula
$AB=BAC$.  The subtlety has apparently been unnoticed before because
't Hooft loops associated with homologically non-trivial cycles such
as $\alpha$ have not been much studied.

The standard definition of the 't Hooft loop is as follows.  We state the
recipe for a general three-manifold $M$ and a circle $\alpha\subset M$.
Let $S=e^{i\phi}$
be a $U(1)$-valued function on $M-\alpha$ (the complement of $\alpha$ in $M$)
that has ``winding number one'' around $\alpha$.  This means  that $ S$ changes
in phase by $2\pi$ in going around a small circle $\beta$ that has linking 
number
one with $\alpha$.  The 't Hooft loop operator is then defined as a gauge
transformation by $S$; under this transformation, one has $a\to a-id\ln S=
 a-d\phi$.

The problem with this definition is that $S$ is described near $\alpha$,
but there is no recipe for what $S$ should look like far away from $\alpha$.
As we will see, when $\alpha$ is homologically nontrivial, a $U(1)$-valued
function $S$ with the claimed properties does not exist.
This problem does not arise in most previous studies of 't Hooft loops
because homologically trivial $\alpha$'s have most often been considered.
(If $\alpha$ is the boundary of an oriented  two-manifold $D\subset M$, 
one can give a recipe for defining $S$ globally with the desired properties,
such that $S=1$ except very near $D$.)  On the other hand, in most studies
of 't Hooft loops, fractional magnetic charge is considered (for the present
case of $U(1)$ gauge theory, this means that $S$ is multivalued in going
around $\alpha$, with the change in phase being a fractional multiple of
$2\pi$).  One then gets interesting properties such as the celebrated
commutation relations of 't Hooft and Wilson loops.  In our present problem,
the electric and magnetic charges are integral, but the cycles are nontrivial.
This will lead to somewhat analogous results.

Before explaining why $S$ does not exist if one expects it to be $U(1)$-valued,
let us first explain in what sense $S$ does exist.  Given any codimension-two
cycle $\alpha$ in a manifold $M$, one can define the Poincar\'e dual
cohomology class $[\alpha]\in H^2(M,\Z)$, and a complex line bundle
${\cal L}$, unique up to isomorphism, with $c_1({\cal L})=[\alpha]$.  Moreover,
${\cal L}$ has a smooth section $s$ with a simple zero along $\alpha$,
of winding number 1 around $\alpha$.  Now, on the complement of $\alpha$,
define $S$ by  $S=s/|s|$.  $S$ has the desired properties $|S|=1$ and winding
number one around $\alpha$, but $S$ is a section of ${\cal L}$ rather than
a $U(1)$-valued function.

At this point, we can readily show the converse: if $[\alpha]\not= 0$, 
then $S$ does not exist
as an ordinary function.  Let $S'$ be a hypothetical $U(1)$-valued function
with winding number one around $\alpha$.  Then $S/S'$ has no winding
number around $\alpha$, and hence extends over $\alpha$ as a smooth and
everywhere nonzero section of ${\cal L}$. Such a function is a trivialization
of ${\cal L}$.  So $S'$ can only exist if
${\cal L}$ is trivial, or in other words if $[\alpha]=0$.

If $[\alpha]\not= 0$ and we define an 't Hooft loop using a ``gauge
transformation'' by $S$, what will we get?  A charged field $\Psi$
will be transformed by this ``gauge transformation'' to $S\Psi$.  If
$\Psi$ is a section of a line bundle ${\cal M}$, then $S\Psi$ is a section
of ${\cal L}\otimes {\cal M}$.  
The operation ${\cal M}\to {\cal L}\otimes {\cal M}$ shifts the
first Chern class of ${\cal M}$ by  $c_1({\cal L})$.
In our problem, the first Chern class
of ${\cal M}$ is understood as fundamental string winding number,
so the ``gauge transformation'' by $S$ shifts that winding number.

For this reason, the 't Hooft loop operator on $\S^3/\Z_3$ does not commute 
with
the elementary string winding number.
If as above we measure the elementary string winding number 
by an operator $A'$ that takes the value $\exp(2\pi ik/3)$ when
the first Chern class is $k$, and define the $D$-string winding number
by a 't Hooft loop operator $B'$ that increases $k$ by 1, then we get
the expected commutation relation $A'B'=B'A'\exp(2\pi i/3)$ for states
with a single wrapped threebrane.

The issue we have investigated is actually relevant to a previous
study \gv\
of threebrane wrapping on $\S^3/\Z_n$.  In that work, it was important
that $n$ states can be made by letting the threebrane absorb $k$
fundamental strings for $k=0,1,\dots, n-1$, and that states with
absorbed $D$-strings should not be counted separately.  The relation
$A'B'=B'A'\exp(2\pi i/n)$ that follows from the above analysis makes
clear why this is so.  If $|k\rangle$ is a state with $k$ absorbed fundamental
strings, then a state with $k'$ absorbed $D$-strings is
$\sum_k\exp(2\pi i kk'/n)|k\rangle$.  One cannot specify both the 
number of absorbed fundamental strings and the number of absorbed
$D$-strings, since the relevant operators do not commute.

The group generated by two operators $A$, $B$ with $AB=BA\exp(2\pi i/3)$
has, up to isomorphism, only one irreducible representation, which is
of dimension three.  From the point of view of the SCFT, the dibaryons
$U^N$, $V^N$, and $W^N$ are three states that transform in this representation.
{}From the point of view of the string theory on $AdS_5\times \S^5/\Z_3$,
the wrapped threebrane, with its three possible flat $U(1)$ connections,
has three ground states that transform in this representation.

\newsec{Strings And Monodromies }

\subsec{Preliminaries}

So far our main focus has been point-like states in  the boundary SCFT
associated with branes that wrap various cycles in the 
internal space $X = \S^5/\Z_3 $.  We have also briefly discussed a few
other types of brane configurations.
What remains is to discuss
membrane-like objects in $AdS_5$ that look like ``strings''
on the boundary. As discussed in section 2.2, these can arise from 
{\it (a)} NS5-branes wrapping a 3-cycle,
  {\it (b)} D5-brane wrapping a 3-cycle, or {\it (c)}
 D3-branes wrapping a 1-cycle in $X$.
  Since the membranes
are of codimension two in $AdS_5$, it is possible to have a monodromy
when a particle is taken around a membrane. As we noted in section 2.2, 
each type of membrane is classified by a $\Z_3$ charge, and a total
of 27 membranes (counting the trivial one) can be constructed.
This suggests that the monodromies  might comprise the nonabelian group
$F$ of 27 elements.

The particles that we will use as test objects  to compute
monodromies are the wrapped onebranes
and threebranes that have already been studied in section 2.
The monodromies will arise from two different effects.  One is
simply that  the wrapped branes that
give membranes and the ones that give test particles have electric and
magnetic couplings to $p$-form gauge fields; these couplings lead to numerical
monodromies.
The other effect, seen in the threebrane-fivebrane system,
is somewhat more exotic and involves a certain brane creation process \whan. 
What happens  here is that when a wrapped threebrane  goes around 
a membrane made by wrapping a fivebrane, it returns to itself with
creation of a string.

\def\IR{{\bf R}}
A useful technical aid in the computation is the following.  
The discussion of monodromies will be purely topological, so $AdS_5$
can be replaced by $\R^5$ (with which it coincides topologically).
The membrane worldvolume $M$ can be taken to be a copy of $\R^3\subset \R^5$.
As for the particle worldline, in computing monodromies one takes it
to be a     circle $C=\S^1$ that winds once around the $\R^3$.  The essential
property of the situation is thus that $M$ and $C$ are linked.
To exhibit this linking neatly, it is convenient to compactify $\R^5$
to $Y= \S^5$, in which case $M$ can be compactified to $\S^3$.  Thus
$M$ and $C$ are respectively copies of $\S^3$ and $\S^1$ in $Y$,
and topologically $M$ and $C$ are linked.  The linking means that 
a manifold $B\subset Y$ with boundary $M$ has intersection number 1 with
$C$, and conversely a manifold $B'\subset Y$ with boundary $C$ has intersection
number 1 with $M$.  

As we noted in section 2.2, if the monodromies around the membranes are to
generate the group $F$, then the central element $C$ of $F$ must
be generated by a membrane of type {\it (c)}.  Since $C$ counts dibaryons
and acts trivially on everything else, we expect to find that the monodromy
in going around a membrane of type {\it (c)} is a factor of $e^{2\pi i/3}$
if the test particle is a threebrane wrapped on $\S^3/\Z_3$, and is
otherwise trivial.  On the other hand, membranes of type {\it (a)} and
{\it (b)} must give monodromies $A$ and $B$.  A monodromy $A$, for example,
assigns a phase $e^{2\pi i/3}$ if the test particle is a wrapped
fundamental string, is trivial if the test particle is a wrapped
$D$-string, and (in view of the action of $A$ and $B$ on dibaryons) is
more interesting and will be described later if the test particle is
a wrapped threebrane.  A monodromy $B$ is similar with $F$-strings and
$D$-strings exchanged.

\subsec{Aharonov-Bohm Effect For Branes}

We consider first the monodromies that arise just from electric and magnetic
couplings of the test particle and the membrane to the same $p$-form
gauge field.  (In view of the discussion in the last paragraph, this
means everything except the parallel transport of a dibaryon state
around a membrane made by wrapping a fivebrane.)  How do electric
and magnetic couplings give  monodromies?  The most elementary
example, which we will generalize,  
is the standard Aharonov-Bohm effect in QED. There we have a topologically
non-trivial field configuration, where on taking a particle with unit
charge in a closed loop $C$ around a magnetic
 source, we pick up a phase given by
$\exp( i \int_C A \cdot dx)$. Basically the phase is measured by the line
integral of the vector potential over the world-line of the particle. There
is a higher dimensional example for branes; a $p$-brane with worldvolume
$V$ coupled
to a $(p+1)$-form gauge field $A$ has a worldvolume interaction 
$\int_VA$.  This will create a monodromy if the $p$-brane is parallel
transported around a suitable magnetic source of $A$.
When we parallel transport a probe (made by wrapping a brane on a cycle
in $\S^5/\Z_3$) around an $AdS_5$ membrane (made by wrapping another
brane on another cycle in $\S^5/\Z_3$), we can get a nontrivial monodromy
by this mechanism if the two branes couple electrically and magnetically
to the same  field.  This will occur in the following cases:

{\it (a$'$)} A membrane of type {\it (a)}, made from a wrapped NS5-brane,
and a probe made from a dual fundamental string wrapped on $\S^1/\Z_3$.

{\it (b$'$)} A membrane of type {\it (b)}, made from a wrapped D5-brane,
and a probe made from a dual $D$-string wrapped on $\S^1/\Z_3$.

{\it (c$'$)} A membrane of type {\it (c)}, made from a wrapped D3-brane,
and a probe made from a dual D3-brane wrapped on $\S^3/\Z_3$.

The  monodromies that we will get from these cases are all
the required monodromies summarized at the end of section 3.1
except for the  more complicated mondromy involving a probe D3-brane
and a membrane of type {\it (a)} or {\it (b)}.  We postpone considering
this last case.

For analyzing the Aharonov-Bohm $c$-number monodromies, we consider for
definiteness case {\it (a$'$)}.  It will be evident that the other
cases are similar.

As explained in section 3.1, we can replace $AdS_5$ by $\S^5$ for the
present purposes.  We thus think of the spacetime as $\S^5\times \S^5/\Z_3$.
We consider a fundamental string whose worldvolume is $V_s=C\times
\S^1/\Z_3$, where $C\subset \S^5$ and $\S^1/\Z_3$ is as usual
a generator of $H_1(\S^5/\Z_3)$.  Likewise, we consider an NS5-brane
with worldvolume $V_m=M\times \S^3/\Z_3$, with $M\subset \S^5$
and $\S^3/\Z_3$ a generator of $H_3(\S^5/\Z_3)$.  As explained in
section 3.1, $C$ and $M$ are a circle and a three-sphere which
are ``linked'' in $\S^5$.  

The fivebrane is a magnetic source of
the Neveu-Schwarz two-form field $B$.  The factor in the path integral
 that will
give the monodromy $T$ is, roughly speaking, 
\eqn\micko{T= \exp\left(i\int_{V_s}B_w\right),}
where $B_w$ is the $B$-field created by the fivebrane.
The reason that this is only roughly the right 
formula is that the $B$-field created by the fivebrane is topologically
nontrivial and so cannot be represented globally by a two-form $B_w$.
A safe way to proceed is to let $Z_s$ be a three-manifold with boundary
$V_s$ and rewrite \micko\ as
\eqn\jicko{T=\exp\left(i\int_{Z_s}F_w\right),}
where $F_w=dB_w$ is the gauge-invariant threeform field created by
the fivebrane.  This is a better formula because $F_w$ is gauge-invariant
and is globally defined.

$F_w$ is determined by the following conditions.
First,
\eqn\jins{dF_w=2\pi \delta(Z_m),}
where $\delta(Z_m)$ is understood as a four-form Poincar\'e dual to
the six-manifold $Z_m$; in what follows, analogous delta functions
will be understood similarly.  Second, $F_w$ should obey the three-form
analog of Maxwell's equations.

If $\S^3/\Z_3$ were a boundary in $\S^5/\Z_3$, say the boundary of
a four-manifold $N$, we could obey \jins\ with $F_w=2\pi\delta(N)$.
This is not actually so.  However, three times $\S^3/\Z_3$ vanishes
in $H_3(\S^5/\Z_3,\Z)$ (since that group is $\Z_3$), so we can find a 
four-manifold $N\subset \S^5/\Z_3$ whose boundary is three copies of 
$\S^3/\Z_3$.  With such an $N$, we can obey \jins\ with 
$F_w=(2\pi/3)\delta(N)$.
The formula for the monodromy is now
\eqn\micko{T=\exp\left((2\pi i/3)\int_{Z_s}\delta(N)\right).}
The integral $\int_{Z_s}\delta(N)$
counts the intersection number of $Z_s$ and $N$.  That
intersection number is 1 modulo 3, 
since $B$ has intersection number 1 with $M$
(as $C$ and $M$ are linked in $\S^5$), and $\S^1/\Z_3$ has 
intersection number 1 modulo 3 with $N$
(as $\S^1/\Z_3$ and $\S^3/\Z_3$ are similarly linked in $\S^5/\Z_3$).
Hence the monodromy is $T=\exp(2\pi i/3)$.

This is the expected monodromy from the discussion at the end of section
3.1.  It reflects the following facts: the monodromy for a membrane of
type {\it (a)} is $A$; the eigenvalue of $A$ for a wrapped fundamental
string is $\exp(2\pi i/3)$.  If we use for the test particle a wrapped
$D$-string, a calculation similar to the above gives a trivial monodromy
around the membrane of type {\it (a)}
(since the $D$-string does not couple to the $B$-field created by
the NS5-brane).  For a test particle consisting of a wrapped threebrane,
additional considerations that we come to shortly are relevant.

The other purely numerical monodromies  -- a membrane of type {\it (b)}
and test particle a wrapped string, or a membrane of type {\it (c)} and
any test particle -- can be treated similarly.  In each case, there is 
a nontrivial monodromy precisely if the test particle is electric-magnetic
dual to the membrane.  

\bigskip\noindent{\it The Remaining Case}

It remains only to analyze the more
elaborate monodromy that arises when a test particle made from a wrapped
threebrane is transported around a membrane
made from a wrapped fivebrane.  For definiteness, we will consider
the case of a membrane of type {\it (b)}, made from a wrapped D5-brane.
We expect the monodromy to equal $B$.

We assume that the threebrane that we use as a test particle is
prepared in an eigenstate of $A$, the operator that equals $\exp(2\pi ik/3)$
for a state with $k$ wrapped fundamental strings.  Since the wrapped
threebrane has $C=\exp(2\pi i/3)$, the relation $AB=BAC$ means that
(if the monodromy is equal to $B$) the wrapped threebrane, when transported
around the membrane,  returns with an extra wrapped fundamental string.

We again consider spacetime to be $\S^5\times \S^5/\Z_3$.
We take the threebrane worldvolume to be $Z_3=C\times \S^3/\Z_3$
and the fivebrane worldvolume to be $Z_5=M\times \S^3/\Z_3$.
$C$ and $M$ are as before a circle and a three-sphere in $\S^5$.
We take the two $\S^3/\Z_3$'s (the second factors in $Z_3$ and $Z_5$)
to be distinct and generic.  They then intersect on a circle $Y'\subset
\S^5/\Z_3$ that is a copy of $\S^1/\Z_3$.

We compare two cases. In case (1), $C$ and $M$ are unlinked in $\S^5$,
and in case (2), which is the real case of interest, they have linking
number 1.  We assume that in case (1), no fundamental strings are present.
As one deforms from case (1) to case (2), $C$ passes through $M$, meeting
it (at some stage) at some point $P\in \S^5$, whereupon $Z_3$ and $Z_5$
meet on the circle $P\times Y'$.  In passing through this intersection
to get to case (2),  a fundamental string is created,
connecting the threebrane to the fivebrane, according to a process
described in \whan.  In the final state,
the  worldvolume of this string 
is (up to homotopy) $Q\times Y'$, with $Q$ a path in $\S^5$
from $C$ to $M$.  This means that, in the monodromy described by case (2),
at some moment in parallel transport about the membrane, the 
wrapped threebrane probe has absorbed an elementary string wrapped on
$Y=\S^1/\Z_3$.  This is the expected monodromy.

It remains to justify the assumption that in case (1), there are no net
fundamental strings connecting the threebrane to the fivebrane.
This is so for the following reason.  A wrapped fundamental string ending
on the threebrane carries electric charge (with respect to the $U(1)$
gauge field on the threebrane worldvolume $Z_3$); 
the total electric charge absorbed
on $Z_3$ must vanish as $Z_3$ is compact.  In case (2), the $B$-field
of the fivebrane makes an extra contribution to the absorbed electric
charge, but in case (1) it does not.  Indeed in case (1), by adding
an exact form to the $B$-field of the fivebrane, one can make this
$B$-field vanish identically near the threebrane.

\subsec{Tying Up A Loose End}

Finally, we would like to tie up a loose end in \wbar.

In that paper, wrapped branes in $AdS_5\times \S^5$ and $AdS_5\times
{\bf RP}^5$ were considered.  Most of them were successfully compared
to boundary conformal field theory.  But there was one case for which
no interpretation was offered --
a threebrane wrapped on a generator of $H_1({\bf RP}^5,\Z)=\Z_2$ to
give a membrane in $AdS_5$, which we will call ${\cal M}$.  
In keeping with what has been seen above,
one would guess that the proper interpretation  of ${\cal M}$ is that
there is a monodromy under transport around ${\cal M}$ consisting of some
$\Z_2$ symmetry $\tau$ of the theory.

The $AdS_5\times {\bf RP}^5$ model depends on $\Z_2$-valued discrete theta
angles $\theta_{NS}$ and $\theta_{RR}$ which were described in \wbar.
The gauge group is $SO(2k)$ (for some $k$)
 if $\theta_{NS}=\theta_{RR}=0$ and otherwise
is  of the form $SO(2k+1)$ or $Sp(N)$.
As we will explain presently, the membrane ${\cal M}$ is stable if and only
if $\theta_{NS}=\theta_{RR}=0$. So $\tau$ should be a discrete symmetry
that exists when the gauge group is $SO(2k)$ but not when it is $SO(2k+1)$
or $Sp(k)$.  There is an obvious candidate for such a discrete symmetry,
namely the outer automorphism of $SO(2k)$ generated by a reflection in
one of the coordinates.  $SO(2k+1)$ and $Sp(N)$ have no such outer 
automorphism.

The outer automorphism of $SO(2k)$ actually played an important role in 
\wbar.  The ``Pfaffian particle'' (constructed by wrapping a threebrane
on a generator of $H_3({\bf RP}^5,\Z)=\Z_2$) is odd under this outer 
automorphism,
and so should have a monodromy $-1$ under parallel transport around
${\cal M}$.  This can be seen by an Aharonov-Bohm effect analogous to what was
explained above.

It remains to explain why ${\cal M}$
 is unstable unless $\theta_{NS}=\theta_{RR}=0$.
A membrane is unstable if it can end on a string, for then it decays by
nucleation of string loops.  In \wbar, strings in $AdS_5$ made by
wrapping fivebranes on fourcycles in ${\bf RP}^5$ were considered.
It was shown that for $\theta_{NS}=\theta_{RR}=0$, one can make
a string by wrapping either an NS5-brane or a D5-brane on an
${\bf RP}^4\subset {\bf RP}^5$.  However, if $(\theta_{NS},\theta_{RR})
\not= (0,0)$, then one or the other kind of string is absent.
This arises as follows.  Consider, for example,  a string made by
wrapping an D5-brane.  Let $[H]$ be the cohomology class of the NS
$B$-field.  On the worldvolume $V_5$ 
of a D5-brane, one requires (in the absence of
threebranes)  \eqn\jdn{\left.[H]\right|_{V_5}=0.}  For a D5-brane wrapped on 
${\bf RP}^4$,
this condition is obeyed if and only if $\theta_{NS}=0$.  If threebranes, 
ending
on $V_5$ in a three-manifold $D$, are included, the condition
\jdn\ becomes
\eqn\hjdn{\left.[H]\right|_{V_5}+[D]=0,}
with $[D]$ the Poincar\'e dual to $D$.  Applied to a D5-brane wrapped on
${\bf RP}^4$, this condition states that $[D]$ must be nonzero, and more
specifically that the string made by wrapping the D5-brane must be
the boundary of a membrane made from a wrapped threebrane; this
is the membrane that we have called ${\cal M}$.  Reading
this statement in reverse, ${\cal M}$ can end on a string made from a wrapped
D5-brane if $\theta_{NS}\not= 0$.  By similar reasoning, ${\cal M}$ can end
on a string made from a wrapped NS5-brane if $\theta_{RR}\not= 0$.
In either case, ${\cal M}$ is unstable.

\vskip 30pt
\centerline{\bf Acknowledgments}
It is pleasure to thank to M.~Krogh, S.~N.~Minwalla, and A.~Mikhailov,
for helpful discussions and N.~Nekrasov for collaboration in
an early stage of the project.  The work of S.G. was supported in part by
grant RFBR No 98-02-16575 and Russian President's grant No 96-15-96939. 
The work of E.W. is supported in part by NSF Grant PHY-9513835 and that of 
M.R. by NSF Grant PHY-9802484.

\newsec{Appendix: Topology of the Lens Spaces}

Here we will, for completeness, compute the cohomology and homology
groups of $X=\S^5/\Z_3$.
Since this space is path connected and orientable,
$H_0 (X, \IZ) = H_5 (X, \IZ) = \IZ$.
Because its universal cover is simply $\phi \colon S^5 \longrightarrow  X$,
the fundamental
group is $\pi_1 (X) = \Z_3$ and therefore $H_1 (X, \IZ)=\Z_3$.

To learn more, we study $X$ by viewing it as a Hopf-like
fibration $\psi \colon X \longrightarrow {\bf CP}^2$. 
Indeed,  the 5-sphere 
\eqn\sphere{ \mid z_1 \mid ^2 + \mid z_2 \mid ^2 + \mid z_3 \mid ^2 =1}
admits a $U(1)$ symmetry
 $z_i\to e^{i\alpha}z_i$.  This commutes with the action of $\Z_3$
on $\S^5$ (which is
obtained by restricting $e^{i\alpha}$ to be a cube root of 1),
and so descends to a $U(1)$ action on $X$.
The quotient $X/U(1)$ is ${\bf CP}^2$.
The cohomology of $X$ can be obtained by a spectral sequence using
this fibration; the computation is described in \bott, p. 244.
The result is that, apart from $H^0(X,\Z)=H^5(X,\Z)=\Z$, the nonzero integral
cohomology groups are $H^2(X,\Z)=H^4(X,\Z)=\Z_3$.
It then follows from the Universal Coefficient Theorem (Corollary 15.14.1 in
\bott) that the nonzero homology groups of $X$, apart from $H_0$ and $H_5$,
are $H_1(X,\Z_3)=H_3(X,\Z_3)=\Z_3$.

In turn, a three-cycle $Y=\S^3/\Z_3$ in $X$ is a lens space itself.
Its cohomology can be computed similarly (using the Hopf fibration over
${\bf CP}^1$), and in particular $H^2(Y,\Z)$, which classifies complex 
line bundles over $Y$, is isomorphic to $\Z_3$.


\listrefs
\end